\documentclass[nojss]{jss}

\usepackage{orcidlink,thumbpdf,lmodern}

\usepackage{framed}
\usepackage{booktabs}
\usepackage{tabularx}
\usepackage{amsmath}
\usepackage{amssymb}
\usepackage{subcaption}
\usepackage{kotex}

\newcolumntype{F}{p{2.8cm}}

\DeclareTextFontCommand{\textib}{%
  \fontseries\bfdefault 
  \itshape
}

\newcommand{\bz}{{\mathbf z}}

\DeclareMathOperator*{\argmin}{argmin}
\DeclareMathOperator*{\argmax}{argmax}

\author{Youngseok Song~\orcidlink{0000-0003-3577-7509}\\West Virginia University
   \And Sofia C. Olhede~\orcidlink{0000-0003-0061-227X}\\EPFL
}
\Plainauthor{Youngseok Song, Sofia C. Olhede
}

\title{\pkg{nethist}: An \proglang{R} Package for Nonparametric Graphon Estimation via Network Histograms}
\Plaintitle{nethist: An R package for Nonparametric Graphon Estimation via Network Histograms}
\Shorttitle{\pkg{nethist}: Nonparametric Graphon Estimation}

\Abstract{
Understanding the generative mechanism of real-world networks is crucial for analyzing connection patterns and making inference from network data. Graphons are widely used to model such mechanisms. Network histogram methods are nonparametric approaches based on blockmodel approximations that provide an intuitive view of network connection structures. However, there is a lack of software packages that construct network histograms. We introduce the \proglang{R} package \pkg{nethist}, which implements network histogram-type graphon estimators within a unified interface. This package is applicable to both single-layer and multilayer networks, and includes graphical summaries for examining both local and global network structures. By providing comprehensive network analysis tools, \pkg{nethist} facilitates understanding of complex systems of interrelated vertices.
}

\Keywords{graphon estimation, network histogram, nonparametric network models, statistical network analysis}
\Plainkeywords{graphon estimation, network histogram, nonparametric network models, statistical network analysis}

\Address{
  Youngseok Song\\
  School of Mathematical Data Science\\
  West Virginia University\\
  Morgantown, WV 26506, USA\\
  E-mail: \email{youngseok.song@mail.wvu.edu}\\
  URL: \url{https://ysong.netlify.app/}\\
  \\
  Sofia C. Olhede\\
  Chair of Statistical Data Science\\
  Department of Mathematics\\
  EPFL\\
  1015 Lausanne, Switzerland\\
  E-mail: \email{sofia.olhede@epfl.ch}
}

\begin{document}




\section[Introduction]{Introduction} \label{sec:intro}

Network analysis has received increasing attention across a wide range of disciplines, as complex systems in nature and human society are often represented as networks.
One of the central goals of network analysis is to characterize network-generating mechanisms. Graphons offer a flexible nonparametric framework to model such mechanisms and serve as a basis for statistical network analysis \citep{bickel2009nonparametric, bollobas_riordan_2009, lovasz2012large, wolfe2013Nonparametric, borgsGraphonsNonparametricMethod2017}. One approach to graphon estimation is the network histogram, a piecewise-constant blockmodel approximation of a graphon. Despite growing literature on network histogram-type graphon estimators, software implementations remain limited.
In this paper, we introduce \pkg{nethist}, a user-friendly \proglang{R} package that provides a unified interface for network histogram-based estimation.

The network histogram partitions vertices into groups of approximately equally size and constructs a piecewise-constant estimator by computing block-wise connection probabilities. When a single-layer network is observed, the vertex partition can be obtained via maximum profile likelihood \citep{olhede2014network} or least squares \citep{gao2015Rateoptimal,kloppOracleInequalitiesNetwork2017}. As an extension of the network histogram, an estimator that reduces variance by merging blocks has also been developed \citep{verdeyme2024Hybrid}. In addition, the multi-network histogram has been introduced to estimate layer-specific graphons in a multilayer network \citep{song2026joint}. Apart from histogram-based methods, several graphon estimators have been proposed. Examples include sort-and-smoothing that applies degree-based vertex sorting and total variation distance smoothing \citep{chan2014Consistent}, low-rank matrix approximation that employs singular value decomposition of an adjacency matrix \citep{chatterjee2015Matrix,xu2018Rates}, and an approach using the common neighbor information of a vertex pair \citep{zhang2017Estimating}. These non-histogram-based methods are implemented in the \proglang{R} packages \pkg{graphon} \citep{You2025graphon} and \pkg{randnet} \citep{randnet}.

Despite this methodological development, software supporting network histogram-based graphon estimation remains limited in the \proglang{R} environment. First, currently available software tools are fragmented across different programming languages.  For example, the single-layer network histogram, originally proposed by \citet{olhede2014network}, is written in \proglang{MATLAB} and \proglang{Julia}  (\pkg{NetworkHistogram.jl} \citep{Dufour2023NetworkHistogram}). The  \proglang{Python} package \pkg{pygraphon} \citep{dufour2023pygraphon} implements the network histogram and the block-merging estimator \citep{verdeyme2024Hybrid}. Such fragmentation makes it difficult for \proglang{R} users to apply these tools within the same programming environment, as well as to integrate them with network analysis packages in \proglang{R}. Furthermore, existing software implementations primarily focus on single-layer networks, limiting their applicability to multilayer network analysis. Importantly, to the best of our knowledge, there is no \proglang{R} package that provides network histograms, which makes histogram-based network analysis difficult to access for \proglang{R} users.

\begin{figure}[!t]
\centering
\includegraphics[width=\textwidth]{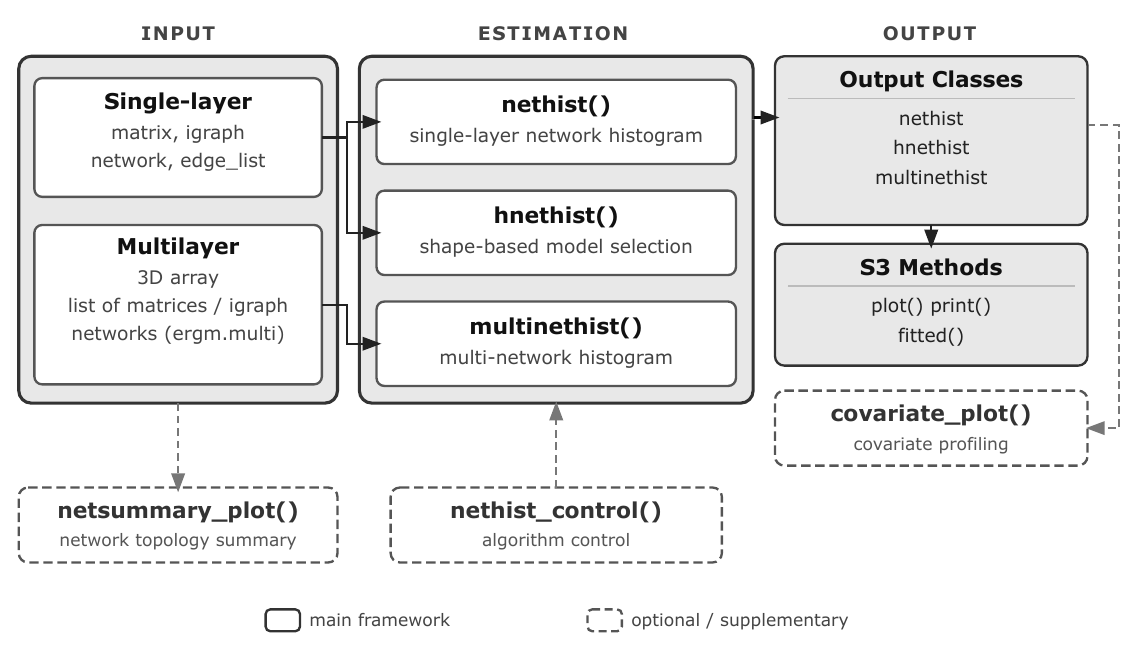}
\caption{\label{fig:overview} Overview of the \pkg{nethist} package.}
\end{figure}

To address these limitations, we develop the \pkg{nethist} \proglang{R} package that constructs network histograms in both single-layer and multilayer network settings \citep{olhede2014network, gao2015Rateoptimal, verdeyme2024Hybrid, song2026joint}. The package provides an integrated workflow, including exploratory data analysis of network structure \citep{maugis2017Topology}, graphon estimation, and visualization of the resulting estimates, as summarized in Figure~\ref{fig:overview}. Built around an S3 class system, it accepts standard network representations as input and returns structured objects that can be summarized and visualized consistently across estimation methods. The S3 design also allows additional estimators to be integrated within the same interface. The main optimization algorithms for constructing network histograms are implemented using \pkg{Rcpp} \citep{eddelbuettel2013Seamless} and \pkg{RcppArmadillo} \citep{eddelbuettel2014RcppArmadillo} for computational efficiency.

The paper is organized as follows. Section~\ref{sec:backgrounds} provides an overview of graphons and network histogram methods. The main and supplementary functions available in the package are introduced in Section~\ref{sec:package}. We demonstrate the applications of the proposed package in Section~\ref{sec:illustrations} using a socio-economic network dataset collected from an Indian village. Section~\ref{sec:summary} concludes the paper and discusses future extensions.

\section{Background} \label{sec:backgrounds}

Section~\ref{subsec:graphon} outlines graphon-based models, and Section~\ref{subsec:nethist} discusses network histogram methods and their extensions.
Throughout the paper, we consider undirected networks with no self-loops or multiple edges.

\subsection{Graphons and related models}
\label{subsec:graphon}

\subsubsection{Scaled graphons for single-layer networks}

A graphon is a symmetric bivariate function $\phi(x,y): [0,1]^2\to [0,1]$, representing the probability of an edge between vertices corresponding to latent positions $x$ and $y$. 
Originally, graphons were introduced to describe limits of sequences of dense graphs \citep{lovasz2012large},
followed by scaled graphons that model sparse networks \citep{bickel2009nonparametric,bollobas_riordan_2009}. 
The key feature of a scaled graphon is the inclusion of a possibly vanishing scaling parameter $\rho_n$, which accommodates sparsity in large networks.
The scaled graphon is defined as
\begin{equation}
\label{def:scaled_graphon}
\phi(x,y) := \rho_n f(x,y).
\end{equation}
Here, $\rho_n \in (0,1)$ represents the sparsity parameter, and $\rho_n f(x,y)\in[0,1]$ for all $(x,y)\in[0,1]^2$, where the symmetric measurable function $f:[0,1]^2 \to [0,\infty)$ satisfies $\iint_{[0,1]^2}f(x,y)dxdy = 1$.
This formulation defines a generative model for a network of size $n$.
Specifically, let the latent variables associated with $n$ vertices, $\{\xi_1,\ldots, \xi_n\}$, be independent and identically distributed samples from $U(0,1)$.
Given $\xi_i$ and $\xi_j$, the edge variable $A_{ij}$ is defined as
\begin{align*}
A_{ij}|\xi_i, \xi_j &\sim \text{Bernoulli}(\rho_n f(\xi_i,\xi_j)), \quad 1\leq i < j \leq n,
\end{align*}
and the remaining entries are defined by $A_{ji}=A_{ij}$.

\subsubsection{Scaled Set of Graphons for Multilayer Networks}

Many real-world networks involve multiple types of relationships among the same entities. 
Each type of relationship forms a distinct layer, offering different perspectives on characteristics of a population. Scaled sets of graphons facilitate joint analysis of all layers while allowing layer-wise heterogeneity in sparsity levels and structural patterns \citep{song2026joint}. 
For each layer $\ell \in [L]$, let $f^{(\ell)}:[0,1]^2 \to [0,\infty)$ be a symmetric measurable function satisfying $\iint_{[0,1]^2} f^{(\ell)}(x,y)dxdy = 1$. 
A scaled set of graphons is defined as
\begin{equation*}
\left\{\phi^{(\ell)}: \phi^{(\ell)}(x,y) = \rho_n^{(\ell)} f^{(\ell)}(x,y), \ell=1,\ldots,L\right\},
\end{equation*}
where $\rho_n^{(\ell)} \in (0,1)$ represents the sparsity parameter of the $\ell$th layer, and $\phi^{(\ell)}(x,y) \in [0,1]$ for all $(x,y)\in[0,1]^2$.
This construction can be used to generate a multilayer network on a common vertex set. All layers use the same latent variables, which are independently drawn from $U(0,1)$ as in the single-layer network. Given the latent variables $\xi_i$ and $\xi_j$, the edge variable of the $\ell$th layer $A_{ij}^{(\ell)}$ is generated according to
\begin{align*}
A_{ij}^{(\ell)}|\xi_i, \xi_j &\sim \text{Bernoulli}(\rho_n^{(\ell)} f^{(\ell)}(\xi_i,\xi_j)), \quad 1\leq i < j \leq n, 1\leq \ell\leq L.
\end{align*}

\subsection{Network Histograms}\label{subsec:nethist}

The network histogram is a nonparametric method for estimating graphons, providing a simple and intuitive view of network connection structure. They partition vertices into groups of equal size with similar connection patterns and then calculate the normalized edge densities for each pair of groups. In these methods, the groups are typically determined based on edge probabilities among vertices. The bandwidth, defined as the group size, is set equal across all groups. The resulting histogram is then a step-function approximation to the function $f$ in \eqref{def:scaled_graphon}. 
This approximation is analogous in that histograms approximate a density function using a step function.

\subsubsection{Network Histogram for Single Layer}

The network histogram for single-layer networks, introduced by \citet{olhede2014network}, is formulated as follows. Let $\bz = (z_1,\ldots, z_n)$ denote the group labels of $n$ vertices, where $z_i \in \{1,\ldots,k\}$. Then, the partition consists of $k-1$ groups each of size $h$, and one group formed by the remaining vertices. The optimal labels $\widehat{\bz}=(\widehat{z}_1,\ldots, \widehat{z}_n)$ are determined by either maximizing profile log-likelihood \citep{olhede2014network},
\begin{equation}
\label{eq:nethist_PLL}
\begin{aligned}
\widehat{\bz}=\argmax_{\bz} \sum_{i<j}\Big\{&A_{ij}\log \overline{A}_{z_i z_j} + (1-A_{ij})\log (1-\overline{A}_{z_i z_j})\Big\},
\end{aligned}
\end{equation} 
or least squares \citep{gao2015Rateoptimal,kloppOracleInequalitiesNetwork2017},
\begin{equation*}
\begin{aligned}
\widehat{\bz}
=\argmin_{\bz} \frac{1}{n^2}\sum_{i,j=1}^n\Big(&A_{ij}- \overline{A}_{z_i z_j}\Big)^2,
\end{aligned}
\end{equation*} 
where $\overline{A}_{z_i z_j}$ is the edge density of the block corresponding to the group pair $(z_i, z_j)$. 
Given the labels $\widehat{\bz}$, the edge density of the $(a,b)$th block is denoted by $\overline{A}_{ab}$ if $\widehat{z}_i=a$ and $\widehat{z}_j=b$. Then, the network histogram is given by
\begin{align}
\label{def:nethist}
\hat{f}_{ab}={\hat{\rho}_n}^{-1}\overline{A}_{ab}, \quad \hat{\rho}_n=\binom{n}{2}^{-1}\sum_{i<j}A_{ij}, \quad 1\leq a,b \leq k,
\end{align}
where $\hat{\rho}_n$ is the empirical edge density.
The bandwidth $h$, which determines the group size in the partition, can be selected by a data-driven procedure that minimizes an upper bound on the mean integrated squared error (MISE) of the graphon estimator $\hat{f}= \{\hat{f}_{ab}\}_{1\leq a,b\leq k}$ \citep{olhede2014network}.

\subsubsection{Hybrid network Histogram with General Shapes}

\citet{verdeyme2024Hybrid} propose the stochastic shape model as an alternative to the stochastic block model. This model constructs shapes by grouping blocks. Using these shapes, the graphon estimator is defined as a step function. Throughout this paper, we refer to this estimator as the hybrid network histogram, following the term ``hybrid approach'' used in their work.

The hybrid network histogram uses the network histogram as an initial fit and then applies a one-dimensional clustering algorithm to merge $\binom{k+1}{2}$ blocks into $s$ clusters, where each cluster is referred to as a shape. Throughout the paper, we call this clustering step the merging step. The number of clusters is chosen to minimize the Bayesian Information Criterion (BIC), which is given by
\begin{equation*}
  \mathrm{BIC} = -2\log \sum_{i<j}\Big\{A_{ij}\log \overline{A}_{z_i z_j, s} + (1-A_{ij})\log (1-\overline{A}_{z_i z_j, s})\Big\} + s \log \frac{n(n-1)}{2},
\end{equation*}
where $\overline{A}_{z_i z_j, s}$ denotes the estimated connection probability for the block indexed by $(z_i, z_j)$ when using $s$ shapes. Given the resulting partition $\{\mathcal{S}_1,\ldots,\mathcal{S}_s\}$, the histogram is calculated as 
\begin{align}
\label{def:hnethist}
\hat{f}_{ab}={\hat{\rho}_n}^{-1}\overline{A}_{\mathcal{S}_r}, \quad (a,b)\in \mathcal{S}_r, \quad 1\leq r\leq s
\end{align}
where $\overline{A}_{\mathcal{S}_r}$ denotes the edge density over blocks in shape $\mathcal{S}_r$. The hybrid network histogram has a smaller variance than the estimator in \eqref{def:nethist}, as it merges blocks with similar connection probabilities.

\subsubsection{Multi-Network Histogram}

The multi-network histogram jointly estimates a scaled set of graphons \citep{song2026joint}. 
When each layer of a multilayer network is defined on a common set of vertices, latent variables associated with vertices are shared across layers and can be used to summarize vertex-specific characteristics. This assumption allows vertices to be grouped using observed edges across multiple layers, enabling finer-resolution graphon estimation even in sparse layers where the underlying structure is difficult to be identified.

Let $\bz=(z_1,\ldots, z_n)$ denote the common group labels across all layers. The estimated labels $\widehat{\bz}$ are obtained by maximizing the joint profile log-likelihood, which decomposes into the sum of layer-wise profile log-likelihoods in \eqref{eq:nethist_PLL},
\begin{equation*}
\begin{aligned}
\widehat{\bz} := \argmax_{\bz}\sum_{\ell=1}^L\sum_{i<j}\Big\{&A_{ij}^{(\ell)}\log \bar{A}_{z_i z_j}^{(\ell)}
+ (1-A_{ij}^{(\ell)})\log (1-\bar{A}_{z_i z_j}^{(\ell)})\Big\},
\end{aligned}
\end{equation*}
where $\overline{A}_{z_i z_j}^{(\ell)}$ is the edge density of the block indexed by the group pair $(z_i,z_j)$ in the $\ell$th layer. 
Given $\widehat{\bz}$, let $\overline{A}_{ab}^{(\ell)}$ denote the edge density of the block corresponding to the group pair $(a,b)$ in the $\ell$th layer. That is, $\overline{A}_{ab}^{(\ell)}=\overline{A}_{\widehat{z}_i \widehat{z}_j}^{(\ell)}$ whenever $\widehat{z}_i=a$ and $\widehat{z}_j=b$.
Then, the multi-network histogram is defined as
\begin{align}
\label{def:multinethist}
\hat{f}_{ab}^{(\ell)}&=\left(\hat{\rho}_n^{(\ell)}\right)^{-1}\overline{A}_{ab}^{(\ell)}, \quad
\widehat{\rho}_n^{(\ell)} = \binom{n}{2}^{-1}\sum_{i<j}A_{ij}^{(\ell)}, \quad \ell=1,\ldots, L,
\end{align}
where $\widehat{\rho}_n^{(\ell)}$ denotes the empirical edge density of the $\ell$th layer. The bandwidth $h$ can be selected using Algorithm 2 of \citet{song2026joint}, which is derived by minimizing an upper bound on the weighted MISE of the graphon estimator $\hat{f}^{(\ell)}= \{\hat{f}_{ab}^{(\ell)}\}_{1\leq a,b\leq k, 1\leq \ell\leq L}$.

In the special case where all layers share the same function $f$ but may have different sparsity levels, a pooled histogram can be constructed using a sparsity-based weighted average of the layer-wise histograms. This method is referred to as a homogeneous multi-network histogram, which can provide higher-resolution estimates.

\subsubsection{Implementation Details}

Constructing the network histograms in \eqref{def:nethist}, \eqref{def:hnethist}, and \eqref{def:multinethist} requires maintaining equal group sizes throughout the optimization process. Existing stochastic block model estimation algorithms, such as the variational expectation-maximization algorithm \citep{daudinMixtureModelRandom2008} and the Gibbs sampler \citep{snijders1997Estimation}, give piecewise-constant representations of edge probabilities, similar to network histograms. However, they do not enforce equal-sized groups and thus cannot be directly applied to construct network histograms.

The package \pkg{nethist} adopts a swap-based greedy search strategy that optimizes the objective function over group labels while preserving equal group sizes. The initial group assignment is obtained using a spectral ordering. The algorithm then repeatedly proposes pairwise exchanges between randomly selected vertices in different groups. A proposed swap is accepted if it improves the objective function. The iterations terminate when no improvement is observed over a fixed number of consecutive steps. The core algorithms are written in \proglang{C++} using \pkg{Rcpp} and \pkg{RcppArmadillo} to address substantial computational overhead due to repeated swap evaluations and incremental matrix updates.

\section{Package Overview} \label{sec:package}

This section provides an overview of the main tools available in the \pkg{nethist} package. We introduce functions for implementing the graphon estimation methods outlined in Section~\ref{sec:backgrounds}, along with their S3 return classes. This section also describes functions for exploring the structure of the observed network and for summarizing the estimated graphon.

\subsection{Graphon Estimation} \label{subsec:estimation}

In the \pkg{nethist} package, graphon estimation is performed by \code{nethist} \citep{olhede2014network}, \code{hnethist} \citep{verdeyme2024Hybrid}, or \code{multinethist} \citep{song2026joint}. The basic usage of these functions is as follows:
\begin{Schunk}
\begin{Sinput}
R> nethist(A, h, method = "PLL", control = nethist_control()) 
R> hnethist(A, h, method = "LSE", control = nethist_control())
R> multinethist(A, h, method = "PLL", common_f, control=nethist_control()) 
\end{Sinput}
\end{Schunk}

\subsubsection{Standard usage}

The functions \code{nethist}, \code{hnethist}, and \code{multinethist} share a common interface, with arguments that define the network input, histogram bandwidth selection, and graphon estimation criterion. The argument \code{A} specifies the network input. For single-layer methods \code{nethist} and \code{hnethist}, the input can be an adjacency matrix that is either dense \code{base::matrix} or sparse \code{Matrix::dgCMatrix}, an edge list that is a two-column matrix or data frame of vertex index pairs, or a network object from the packages \pkg{igraph} \citep{igraph2006} or \pkg{network} \citep{network2008}. 
For multilayer networks, \code{multinethist} takes \code{A} as a collection of network layers defined on a common set of vertices. It can be provided as a three-dimensional \code{array}, where each slice corresponds to a layer, a list of \code{igraph::igraph} objects sharing the same vertices, or an \code{ergm.multi::combined_network} object~\citep{ergm_mutli}. The argument \code{h} controls the network histogram bandwidth. By default, the data-driven selection procedures discussed in Section~\ref{subsec:nethist} are applied. Alternatively, users may specify a histogram bandwidth ranging from two to the number of vertices. The third argument \code{method} specifies an optimization criterion, either maximum profile log-likelihood (\code{method = "PLL"}) or least squares (\code{method = "LSE"}). Following the corresponding original papers, the functions \code{nethist} and \code{multinethist} use \code{PLL} by default, and the function \code{hnethist} uses \code{LSE} by default. The argument \code{common\_f}, only available in \code{multinethist}, determines whether a homogeneous multi-network histogram is fitted (\code{FALSE} by default).

\subsubsection{Additional tuning}

The \code{control} argument in the graphon estimation functions accepts a \code{nethist_control} object,
which collects arguments governing the optimization algorithm. This object currently contains the following fields and is designed to accommodate future extensions.

\begin{enumerate}
\item \code{algorithm}: Optimization algorithm. Currently only \code{``greedy"} is implemented.
\item \code{max\_itr}: Maximum number of iterations. Default is $5 \times 10^6$.
\item \code{greedy\_swap\_rule}: vertex-pair selection rule for \code{algorithm = "greedy"}. At present, only \code{"single\_random"} is supported, which draws one pair of vertices uniformly at random and swaps their block labels upon improvement of the objective function corresponding to the selected \code{method}.
\item \code{greedy\_stop\_threshold}: Integer controlling early stopping for \code{algorithm = "greedy"}. The algorithm stops when the objective function has not improved for a specified number of consecutive iterations. Default is 20{,}000.
\item \code{verbose}: Logical indicating whether to print progress messages during fitting. Default is \code{FALSE}.
\end{enumerate}

\subsubsection[Return value: Classes nethist, multinethist, hnethist]{Return value: Classes \code{nethist}, \code{multinethist}, \code{hnethist}}\label{subsubsec:return}

\begin{table}[t]
\centering
\caption{Fields of \texttt{nethist} and \texttt{multinethist} objects. The number of vertices, blocks, and layers are denoted by $n$, $k$, and $L$, respectively.}
\label{tab:nethist-multinethist-fields}
\begin{tabularx}{\textwidth}{F|X}
\hline
Field & Description  \\
\hline
\code{cluster}
  & Integer vector of length $n$ representing vertex group labels $\widehat{\bz}=(\widehat{z}_1,\ldots, \widehat{z}_n)$ where $\widehat{z}_i \in \{1,\ldots,k\}$\\
\code{thetahat}
  & $k\times k$ estimated probability matrix, $\widehat{\Theta}=\{\overline{A}_{ab}\}_{a,b=1,\ldots, k}$, for \code{nethist} ($k\times k\times L$ array, $\widehat{\Theta}=\{\overline{A}_{ab}^{(\ell)}\}_{a,b=1,\ldots, k; \ell=1,\ldots, L}$, for \code{multinethist}) \\
\code{rho\_hat}
  & Edge density estimate $\hat{\rho}_n$ for \code{nethist} (length $L$ vector $\hat{\rho}_{n}^{(\ell)}$ for \code{multinethist}) \\
\code{h} & Integer histogram bandwidth\\
\code{normalized\_LL}
  & Log-likelihood of the fitted model, normalized by total observed edges (summed over layers for \code{multinethist})\\
\code{MSE}
  & Mean squared error of the fitted model (summed over layers for \code{multinethist}) \\
\code{method}
  & Loss function used, either \code{PLL} or \code{LSE} \\
\code{homogeneous}
  & Logical (for \code{multinethist} only) indicating whether a common graphon $f$ is assumed across all layers when \code{TRUE}.\\
\hline
\end{tabularx}
\end{table}

\begin{table}[!t]
\centering
\caption{Fields of objects of class \code{hnethist}. The number of vertices and blocks are denoted by $n$ and $k$, respectively.}
\label{tab:hnethist-fields}
\begin{tabularx}{\textwidth}{F|X}
\hline
Field & Description \\
\hline
\code{cluster}
  & Integer vector of length $n$ representing vertex group labels $\widehat{\bz}=(\widehat{z}_1,\ldots, \widehat{z}_n)$ where $\widehat{z}_i \in \{1,\ldots,k\}$\\
\code{thetahat}
  &  Estimated probability matrix of the selected model obtained from the merging step  \\
\code{rho\_hat}
  &  Edge density estimate $\hat{\rho}_n$ \\
\code{h}  & Integer histogram bandwidth\\
\code{normalized\_LL}
  & Log-likelihood of the selected model, normalized by observed edges \\
\code{MSE}
  & Mean squared error of the selected model\\
\code{method}
  & Loss function used, either \code{PLL} or \code{LSE} \\
\code{blockcluster}
  & an object from a clustering method for merging the $\binom{k+1}{2}$ unique entries of \code{\$thetahat} (upper triangle including diagonal) into $s$ shapes \\
\code{BIC}
  & BIC value of the selected model \\
\code{s}
  & Number of distinct connection-probability shapes \\
\code{details}
  & List of at most $\binom{k+1}{2}$ candidate models, one per $s$ \\
\code{initial}
  & \code{nethist} object; initial estimate used as the starting point for the merging step \\
\hline
\end{tabularx}
\end{table}

The functions \code{nethist}, \code{multinethist}, and \code{hnethist} return S3 objects whose classes correspond to the function names. 

The \code{nethist} class provides a structure that stores the components used in constructing a network histogram, as summarized in Table~\ref{tab:nethist-multinethist-fields}.
The field \code{\$rho\_hat} indicates the edge density of the input network. The entry \code{\$thetahat[a,b]} corresponds to the estimated connection probability between a vertex in group $a$ and one in group $b$. The group memberships are stored in \code{$cluster}, and the histogram bandwidth is given by \code{\$h}. The \code{\$method} field records the optimization criterion used for graphon estimation (\code{"PLL"} or \code{"LSE"}). The fields \code{\$normalized\_LL} and \code{\$MSE} store the normalized log-likelihood and mean squared error, respectively.

The \code{hnethist} class extends the \code{nethist} class  and contains additional fields for hybrid network histograms, as summarized in Table~\ref{tab:hnethist-fields}. The \code{nethist} fit used as the starting point of the merging step is stored in \code{\$initial}. The values of \code{\$cluster}, \code{\$rho\_hat}, and \code{\$h} in the \code{hnethist} object are identical to those in \code{\$initial}.
The fields \code{\$s} and \code{\$BIC} record the number of shapes and the BIC values, respectively. The block clustering results used for shape construction are stored in \code{\$blockcluster}. The full model search path is stored in \code{\$details}. 
The fields \code{\$thetahat}, \code{\$normalized\_LL}, and \code{\$MSE} summarize the estimation results obtained after the merging step.

The \code{multinethist} class is the multilayer counterpart of the \code{nethist} class, as shown in Table~\ref{tab:nethist-multinethist-fields}. The \code{\$rho\_hat} field changes from a scalar to a vector, where the $l$th entry represents the edge density of the $l$th layer. The estimated \code{\$thetahat} is extended to a three-dimensional array, and the entry \code{\$thetahat[a,b,l]} gives the connection probability between a vertex in group $a$ and one in group $b$ on the $l$th layer. The logical field \code{\$homogeneous} indicates whether a homogeneous multi-network histogram is fitted.

The network histogram can be obtained by normalizing \code{\$thetahat} with \code{\$rho\_hat}. For \code{nethist} in \eqref{def:nethist} and \code{hnethist} in \eqref{def:hnethist}, this is given by \code{\$thetahat / \$rho\_hat}. For \code{multinethist} in \eqref{def:multinethist}, layer-wise estimates are computed as \code{\$thetahat[,,l] / \$rho\_hat[l]} for each $l$. In the next section, we introduce functions that can directly compute and visualize the network histogram.

\subsection{Object Methods and Visualization} \label{subsec:vis_ftns}

The \pkg{nethist} package provides utility functions for exploratory and post-fitting analysis.

\subsubsection{S3 Methods for Return Values}

The package provides S3 methods for the \code{nethist}, \code{hnethist}, and \code{multinethist} classes, including \code{print}, \code{plot}, and \code{fitted}.

The \code{print} method provides a standard way to view the estimation results. 
Users may call it either by typing the object name or by using \code{print()} as follows:
\begin{Schunk}
\begin{Sinput}
R> x #if x is a nethist, hnethist, or multinethist object
R> print(x, ...) #... passed to base::print.default()
\end{Sinput}
\end{Schunk}
The \code{print} method displays the estimated probability matrix or array (\code{$thetahat}), the optimization criterion used for estimation (\code{\$method}), and either the normalized log-likelihood (\code{$normalized_LL}) or MSE (\code{$MSE}), depending on the estimation method. For \code{hnethist} objects, it additionally reports the BIC value and the number of shapes in the selected model. It also prints a summary of accessible fields in the returned object.

The estimated network structure can be visualized via the \code{plot} method as follows:
\begin{Schunk}
\begin{Sinput}
R> plot(x, type, idx_order, colorkey, power, prob) 
\end{Sinput}
\end{Schunk}
The \code{type} argument specifies the type of plot to be generated. For all classes, \code{type = "nethist"} produces a network histogram plot, and \code{type = "prob"} produces a probability matrix plot. For \code{hnethist}, \code{type = "BIC"} provides a plot of BIC values against the number of shapes. 
For \code{type = "nethist"} and \code{type = "prob"}, the \code{idx\_order} argument specifies the ordering of groups (\code{1:max(object\$cluster)} by default). 
The \code{colorkey} argument controls whether the color key is displayed (\code{FALSE} by default). The \code{power} argument applies a power transformation to the estimated values before mapping them to the color scale when \code{type = "nethist"} (0.25 by default). The logical argument \code{prob} determines whether the estimated probabilities are displayed in each cell when \code{type = "prob"} (\code{FALSE} by default).

For specified subsets of vertex pairs, the \code{fitted} method returns entries of the network histogram or the estimated edge probability matrix. For multilayer networks, a subset of layers may also be specified. The method can be used as follows:
\begin{Schunk}
\begin{Sinput}
R> fitted(x, set1, set2, type)        # nethist or hnethist object
R> fitted(x, set1, set2, layer, type) # multinethist object
\end{Sinput}
\end{Schunk}
The \code{set1} and \code{set2} arguments define vertex subsets as integer vectors. If \code{NULL}, all vertices are used, though a warning is issued when the resulting matrix exceeds $3000 \times 3000$ entries. The \code{layer} argument selects a subset of layers as an integer vector and is only available for \code{multinethist} objects. The \code{type} argument controls whether the network histogram (\code{"nethist"} by default) or the edge probability matrix (\code{"prob"}) is returned.

\subsubsection{Supplementary Visualization Tools}

The package provides two supplementary plotting functions, one summarizing network statistics and the other displaying covariate distributions within each group.

The \code{netsummary_plot} function creates the violin plots of 2-star and $k$-cycle statistics across a range of vertex subsample sizes. This plot, introduced by \citet{maugis2017Topology}, provides a graphical summary of local network structure, including triadic closure, connectivity patterns, and structural heterogeneity. It facilitates comparisons of structural properties across networks.
The function can be used as follows:
\begin{Schunk}
\begin{Sinput}
R> netsummary_plot(A, subsample_sizes, max_cycle_order, ...)
\end{Sinput}
\end{Schunk}
The argument \code{A} can be an adjacency matrix, an \code{igraph} object, or a \code{network} object. The \code{subsample\_sizes} argument specifies the vertex subsample sizes. By default, the function uses the data-driven subsample size selection procedure proposed in Algorithm 2 of \citet{maugis2017Topology}. The \code{max\_cycle\_order} specifies the maximum cycle order and must be an integer between 3 and 7 (4 by default). 

The \code{covariate_plot} function plots the distribution of vertex covariates in each group. This plot provides a graphical summary of group-specific covariate patterns, where the group membership is determined by the \code{$cluster} field in the \code{nethist}, \code{hnethist}, and \code{multinethist} objects. 
\begin{Schunk}
\begin{Sinput}
R> covariate_plot(object, covariate, idx_order)  
\end{Sinput}
\end{Schunk}
The \code{object} argument is a \code{nethist}, \code{hnethist}, or \code{multinethist} object. The \code{covariate} argument is a vertex-level covariate vector, which can be either categorical (\code{factor}) or numeric. If \code{covariate} is categorical, a stacked bar chart is drawn for each group. If it is numeric, a violin plot is drawn instead. The \code{idx\_order} argument reorders the groups along the horizontal axis (\code{1:max(object$cluster)} by default).

\section{Illustrations: Indian Household Socio-Economic Networks} \label{sec:illustrations}

\begin{table}[!t]
\centering
\caption{Layers in the \code{IndianVil} dataset. Each layer represents one type of socio-economic relationship among households.}
\label{tab:layer_info}
\begin{tabular}{lll}
\hline
Layer No. & Relationship & Description \\
\hline
1  & Borrow money     & borrows money from another household \\
2  & Give advice      & gives advice to another household \\
3  & Help decision    & helps another household make decisions \\
4  & Kero rice come   & receives kerosene or rice from another household \\
5  & Kero rice go     & provides kerosene or rice to another household \\
6  & Lend money       & lends money to another household \\
7  & Medic            & provides medical help to another household \\
8  & Nonrel           & Non-relative relationship between households \\
9  & Rel              & Kinship or relative relationship between households \\
10 & Temple company   & accompanies another to the temple \\
11 & Visit come       & receives visits from another household \\
12 & Visit go         & visits another household \\
\hline
\end{tabular}
\end{table}

This section illustrates the use of the tools in \pkg{nethist} with the \code{IndianVil} dataset provided in the package. The dataset is a subset of the original network data from a study of 75 villages in Karnataka, India \citep{banerjee2013Diffusion,indianvildata}. It consists of networks from one of the villages and has also been used as an illustrative example in \citet{song2026joint}. Before including the dataset in the package, vertices with zero degree across all layers were removed, and each layer was converted into a simple undirected network. The resulting dataset is a 231 $\times$ 231 $\times$ 12 array, with a total of 4506 edges. Each layer represents a distinct type of socio-economic relationship among 231 households. The dataset contains 12 relationship types, summarized in Table~\ref{tab:layer_info}. These relationship types appear as the names of the third array dimension, as follows. 
\begin{Schunk}
\begin{Sinput}
R> library("nethist")
R> data(IndianVil)
R> layer_spec <- dimnames(IndianVil)[[3]]
R> layer_spec
\end{Sinput}
\begin{Soutput}
 [1] "borrow money"   "give advice"    "help decision" 
 [4] "kero rice come" "kero rice go"   "lend money"    
 [7] "medic"          "nonrel"         "rel"           
[10] "temple company" "visit come"     "visit go"      
\end{Soutput}
\end{Schunk}

\begin{figure}[!t]
\centering
\includegraphics{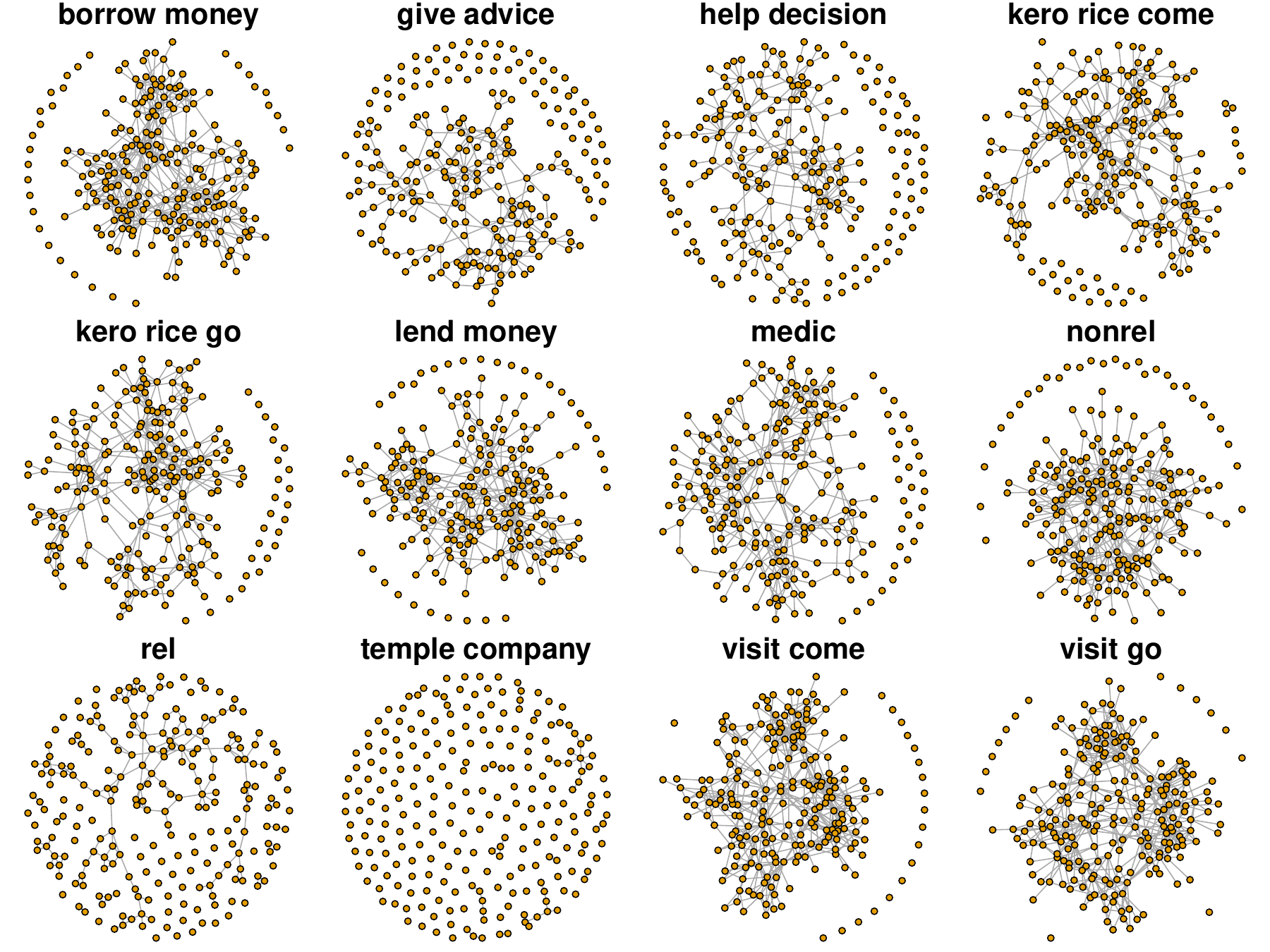}
\caption{\label{fig:IndianVil} \label{fig:igraph_indianvil} \pkg{igraph} plots of the layers in the \code{IndianVil} dataset included in \pkg{nethist}. Each panel represents one layer with vertices corresponding to households and edges indicating reported ties.}
\end{figure}

The following code was used to visualize the dataset using the \pkg{igraph} package. 
\begin{Schunk}
\begin{Sinput}
R> library("igraph")
R> set.seed(2026)
R> par(mfrow = c(3, 4), mar = c(0, 0, 2, 0))
R> for (k in 1:12) {
+    g <- graph_from_adjacency_matrix(IndianVil[, , k], 
+                                     mode = "undirected")
+    plot(g, vertex.size = 5, vertex.label = NA)
+    title(layer_spec[k], cex.main = 2.5)
+  }
\end{Sinput}
\end{Schunk}
Figure~\ref{fig:IndianVil} displays 12 layers in the network with varying levels of sparsity and structural patterns across different socio-economic relationships. The temple company layer is the sparsest, with few inter-household connections and many isolated vertices, while the visit come layer exhibits interactions among many households. The rel layer has a tree-like structure, whereas the nonrel layer contains more connections and exhibits a more complex structure.

\begin{figure}[!t]
\centering
\begin{subfigure}{0.49\textwidth}
\centering
\caption{temple company}
\includegraphics{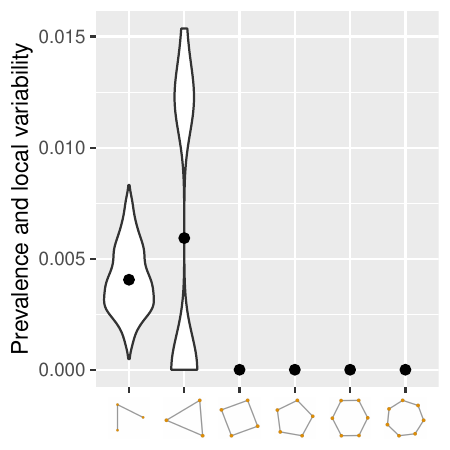}
\end{subfigure}
\begin{subfigure}{0.49\textwidth}
\centering
\caption{visit come}
\includegraphics{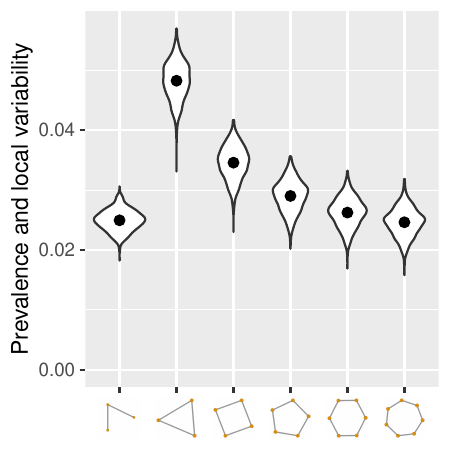}
\end{subfigure}

\vspace{0.5em}

\begin{subfigure}{0.49\textwidth}
\centering
\caption{rel}
\includegraphics{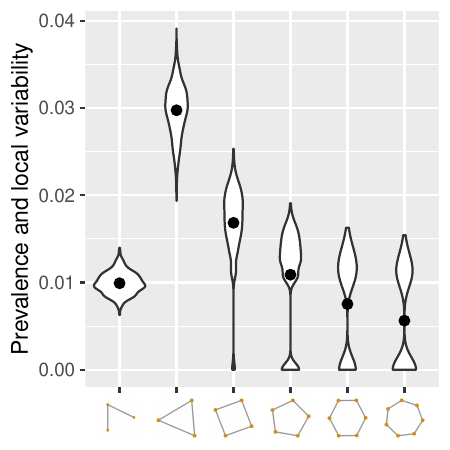}
\end{subfigure}
\begin{subfigure}{0.49\textwidth}
\centering
\caption{nonrel}
\includegraphics{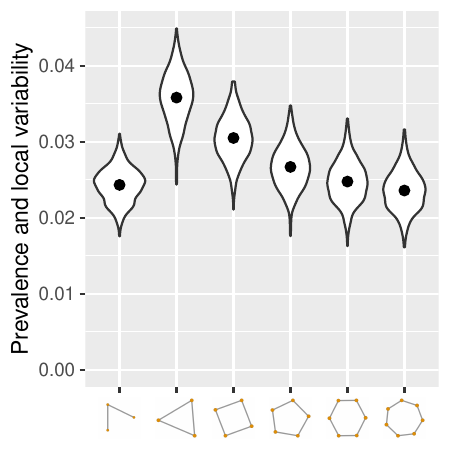}
\end{subfigure}
\caption{\label{fig:IndianVil_netsummary} Network summary plots of the four selected layers. In each subfigure, from left to right: normalized subgraph statistics for 2-stars (relative to triplets with at least one edge) and cycles (triangles, squares, pentagons, hexagons, and heptagons), each scaled by the edge density raised to the number of edges in the corresponding subgraph.}
\end{figure}
To further examine layer-wise characteristics, \code{netsummary\_plot()} is applied to the four selected layers: temple company, visit come, rel, and nonrel.
\begin{Schunk}
\begin{Sinput}
R> set.seed(2026) #For reproducibility
R> par(mfrow=c(1, 4), mar=c(1, 1, 3, 1))
R> for (k in c(10, 11, 9, 8)) {
+    netsummary_plot(IndianVil[,,k], max_cycle_order = 7, 
+                    subsample_sizes = 150)
+  }
\end{Sinput}
\end{Schunk}
Figure~\ref{fig:IndianVil_netsummary} displays the normalized 2-star and $k$-cycle ($k=3,\ldots,7$) statistics of the four selected layers. These layers exhibit a strong tendency toward triadic closure, as normalized triangle statistics are larger than 2-stars across all layers. Each socio-economic relationship also displays distinct interaction patterns. The temple company layer in Figure~\ref{fig:IndianVil_netsummary}(a) has nonzero statistics only for 2-stars and triangles, due to its extremely sparse connections. Compared with the temple company layer, the visit come layer in Figure~\ref{fig:IndianVil_netsummary}(b) shows higher values across both 2-star and cycle statistics, indicating a denser network structure in the layer. The rel layer in Figure~\ref{fig:IndianVil_netsummary}(c) exhibits larger variation in cycle statistics, suggesting greater diversity in local network structure. However, the nonrel layer in Figure~\ref{fig:IndianVil_netsummary}(d) has less variation in cycle statistics, indicating a more homogeneous local structure than the rel layer.

Following the exploratory data analysis above, we illustrate graphon estimation using the main functions avaialable in the package. We first consider a single-layer example in Section~\ref{subsec:nethist_sl}, followed by a multilayer example in Section~\ref{subsec:multilayer}.

\subsection{Case 1: Single Layer Networks} \label{subsec:nethist_sl}

We focus on the borrow money layer of the \code{IndianVil} dataset and illustrate graphon estimation for this layer.

We first use the \code{nethist()} function. The code below illustrates how to apply the profile log-likelihood estimator (default), the least squares alternative, and how to specify a bandwidth using the \code{h} argument.
\begin{Schunk}
\begin{Sinput}
R> BM_layer <- IndianVil[,,1]
R> set.seed(42); nh_bm_PLL <- nethist(BM_layer) #method = "PLL" by default
R> set.seed(42); nh_bm_LSE <- nethist(BM_layer, method = "LSE")
R> set.seed(42); nh_bm_PLL_h25 <- nethist(BM_layer, h = 25)
\end{Sinput}
\end{Schunk}
The function \code{nethist()} returns a \code{nethist} object, as detailed in Section~\ref{subsec:estimation}, for which a \code{print} method is provided.
The following code prints the \code{nethist} object from the PLL fit. We set \code{digits=2} to save space.
\begin{Schunk}
\begin{Sinput}
R> print(nh_bm_PLL, digits = 2) #Print thetahat upto 2 significant digits
\end{Sinput}
\begin{Soutput}
thetahat:
        [,1]   [,2]   [,3]   [,4]  [,5]   [,6]  [,7]  [,8]   [,9]
 [1,] 0.1268 0.0556 0.0312 0.0816 0.047 0.0035 0.014 0.043 0.0096
 [2,] 0.0556 0.0833 0.0052 0.0104 0.000 0.0174 0.054 0.000 0.0000
 [3,] 0.0312 0.0052 0.1413 0.0156 0.000 0.0069 0.000 0.012 0.0000
 [4,] 0.0816 0.0104 0.0156 0.0072 0.073 0.0156 0.033 0.000 0.0000
 [5,] 0.0469 0.0000 0.0000 0.0729 0.069 0.0000 0.000 0.000 0.0000
 [6,] 0.0035 0.0174 0.0069 0.0156 0.000 0.0942 0.000 0.000 0.0000
 [7,] 0.0139 0.0538 0.0000 0.0330 0.000 0.0000 0.000 0.014 0.0000
 [8,] 0.0434 0.0000 0.0122 0.0000 0.000 0.0000 0.014 0.000 0.0000
 [9,] 0.0096 0.0000 0.0000 0.0000 0.000 0.0000 0.000 0.000 0.0000

Method: Profile Likelihood

normalized likelihood:
-3.91760503088951

Available components:

[1] "cluster"       "thetahat"      "rho_hat"       "normalized_LL"
[5] "MSE"           "method"        "h"            
\end{Soutput}
\end{Schunk}
As mentioned in Section~\ref{subsec:vis_ftns}, the \code{print} method displays \code{\$thetahat}, \code{\$method}, and either \code{\$normalized\_LL} or \code{\$MSE}. The \code{\$thetahat} field shows the estimated $9 \times 9$ connection probability matrix, and the corresponding normalized log-likelihood value is given by -3.9176.

Users may access additional fields that are not displayed by the \code{print} method.
For example, the \code{$rho_hat} field indicates that the edge density of the borrow money layer is 1.73\%. 
\begin{Schunk}
\begin{Sinput}
R> nh_bm_PLL$rho_hat
\end{Sinput}
\begin{Soutput}
[1] 0.01731602
\end{Soutput}
\end{Schunk}
The \code{\$h} and \code{\$cluster} fields store the histogram bandwidth and group assignments, respectively. 
The following code reports these results under both the data-driven and user-specified settings.
\begin{Schunk}
\begin{Sinput}
R> nh_bm_PLL$h #data-driven bandwidth is 24.
\end{Sinput}
\begin{Soutput}
[1] 24
\end{Soutput}
\begin{Sinput}
R> table(nh_bm_PLL$cluster) #group assignments of vertices, data-driven
\end{Sinput}
\begin{Soutput}
 1  2  3  4  5  6  7  8  9 
24 24 24 24 24 24 24 24 39 
\end{Soutput}
\begin{Sinput}
R> nh_bm_PLL_h25$h #user-specified bandwidth is 25.
\end{Sinput}
\begin{Soutput}
[1] 25
\end{Soutput}
\begin{Sinput}
R> table(nh_bm_PLL_h25$cluster) #user-specified
\end{Sinput}
\begin{Soutput}
 1  2  3  4  5  6  7  8  9 
25 25 25 25 25 25 25 25 31 
\end{Soutput}
\end{Schunk}
Under the data-driven bandwidth method, the 231 vertices are partitioned into nine groups, consisting of eight groups with 24 vertices each and one group with 39 vertices.
With a user-specified bandwidth of 25, the resulting partition consists of eight groups with 25 vertices each and one group with 31 vertices.

\begin{figure}[t!]
\centering
\begin{subfigure}{0.49\textwidth}
\centering
\includegraphics{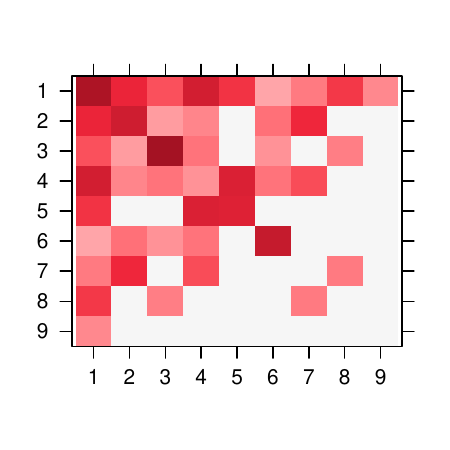}
\caption{}
\end{subfigure}
\begin{subfigure}{0.49\textwidth}
\centering
\includegraphics{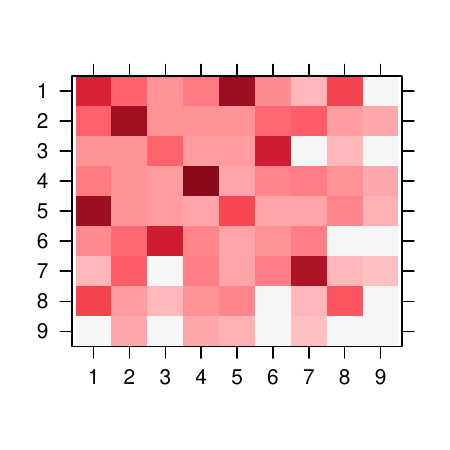}
\caption{}
\end{subfigure}

\begin{subfigure}{0.49\textwidth}
\centering
\includegraphics{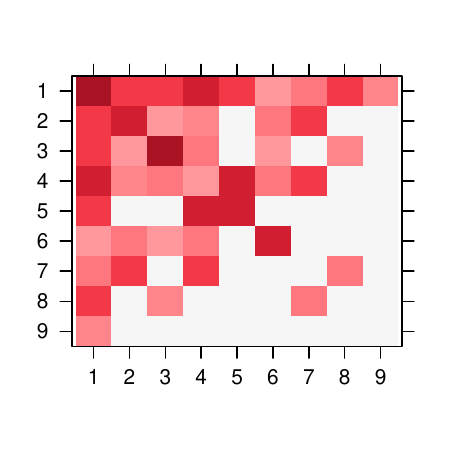}
\caption{}
\end{subfigure}
\begin{subfigure}{0.49\textwidth}
\centering
\includegraphics{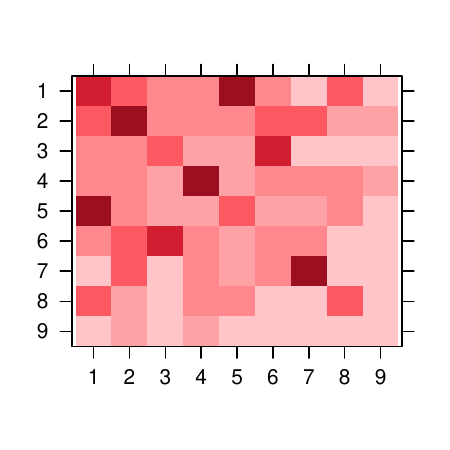}
\caption{}
\end{subfigure}
\caption{Network histograms of the borrow money layer in the \code{IndianVil} dataset. The displayed values represent $\hat{f}^{1/4}$ for better visualization. (a) Network histogram with profile likelihood. (b) Network histogram with least squares. (c) Hybrid network histogram with profile likelihood. (d) Hybrid network histogram with least squares. All panels use the same color scale to facilitate comparison.}
\label{fig:nethist_borrowmoney}
\end{figure}
The \code{nethist} objects are visualized using the \code{plot} method as follows.
\begin{Schunk}
\begin{Sinput}
R> plot(nh_bm_PLL) # Figure 3(a)
R> plot(nh_bm_LSE) # Figure 3(b)
\end{Sinput}
\end{Schunk}
Figures~\ref{fig:nethist_borrowmoney}(a) and (b) present network histograms of the PLL and LSE fits, respectively. A power of 0.25, which is the default value, is used to enhance the clarity of patterns in the plot.
The PLL fit exhibits several near-zero blocks in its network histogram, while the LSE fit displays a relatively smoother network histogram. The two estimates differ due to different group assignments, resulting from distinct optimization criteria. The adjusted Rand index \citep{hubert1985Comparing} between the two partitions is 0.2329, indicating relatively low agreement. The index is computed using the \pkg{mclust} package \citep{mclust}.
\begin{Schunk}
\begin{Sinput}
R> mclust::adjustedRandIndex(nh_bm_PLL$cluster, nh_bm_LSE$cluster)
\end{Sinput}
\begin{Soutput}
[1] 0.2329198
\end{Soutput}
\end{Schunk}

The following code returns entries of the network histogram and the edge probability matrix for vertices 1, 2, and 3 using the \code{fitted} method:
\begin{Schunk}
\begin{Sinput}
R> fit_nh <- fitted(nh_bm_PLL, set1 = c(1,2,3), set2 = c(1,2,3),
+                   type = "nethist")
R> fit_nh
\end{Sinput}
\begin{Soutput}
          [,1]      [,2]     [,3]
[1,] 8.1603261 0.7018229 1.804688
[2,] 0.7018229 0.0000000 2.506510
[3,] 1.8046875 2.5065104 7.323370
\end{Soutput}
\begin{Sinput}
R> fit_prob <-fitted(nh_bm_PLL, set1 = c(1,2,3), set2 = c(1,2,3),
+                    type = "prob")
R> fit_prob
\end{Sinput}
\begin{Soutput}
           [,1]       [,2]       [,3]
[1,] 0.14130435 0.01215278 0.03125000
[2,] 0.01215278 0.00000000 0.04340278
[3,] 0.03125000 0.04340278 0.12681159
\end{Soutput}
\end{Schunk}
The corresponding vertex group indices are 3, 8, and 1, respectively.
\begin{Schunk}
\begin{Sinput}
R> v_grp_idx <- nh_bm_PLL$cluster[c(1,2,3)]
R> v_grp_idx 
\end{Sinput}
\begin{Soutput}
[1] 3 8 1
\end{Soutput}
\end{Schunk}
The edge probability matrix can be obtained directly from \code{\$thetahat} using the corresponding group labels.

\begin{Schunk}
\begin{Sinput}
R> identical(fit_prob,nh_bm_PLL$thetahat[v_grp_idx, v_grp_idx])
\end{Sinput}
\begin{Soutput}
[1] TRUE
\end{Soutput}
\end{Schunk}
The network histogram entries can be calculated by normalizing the edge probability matrix with the edge density \code{\$rho\_hat}.
\begin{Schunk}
\begin{Sinput}
R> identical(fit_nh, fit_prob/nh_bm_PLL$rho_hat)
\end{Sinput}
\begin{Soutput}
[1] TRUE
\end{Soutput}
\end{Schunk}

We now apply \code{hnethist}. As described in Section~\ref{subsec:nethist}, the hybrid network histogram approach aims to reduce variance by smoothing blocks with similar connection probabilities. The code below illustrates both the least squares (default) and profile log-likelihood estimators.
\begin{Schunk}
\begin{Sinput}
R> set.seed(42); hnh_bm_LSE <- hnethist(BM_layer, method = "LSE") #Default
R> set.seed(42); hnh_bm_PLL <- hnethist(BM_layer, method = "PLL")
\end{Sinput}
\end{Schunk}
The following code displays the results of the LSE fit from \code{hnethist}.
\begin{Schunk}
\begin{Sinput}
R> print(hnh_bm_LSE, digits = 2)
\end{Sinput}
\begin{Soutput}
Theta_hat:
        [,1]   [,2]   [,3]   [,4]   [,5]   [,6]   [,7]   [,8]   [,9]
 [1,] 0.0779 0.0283 0.0091 0.0091 0.1537 0.0091 0.0007 0.0283 0.0007
 [2,] 0.0283 0.1537 0.0091 0.0091 0.0091 0.0283 0.0283 0.0041 0.0041
 [3,] 0.0091 0.0091 0.0283 0.0041 0.0041 0.0779 0.0007 0.0007 0.0007
 [4,] 0.0091 0.0091 0.0041 0.1537 0.0041 0.0091 0.0091 0.0091 0.0041
 [5,] 0.1537 0.0091 0.0041 0.0041 0.0283 0.0041 0.0041 0.0091 0.0007
 [6,] 0.0091 0.0283 0.0779 0.0091 0.0041 0.0091 0.0091 0.0007 0.0007
 [7,] 0.0007 0.0283 0.0007 0.0091 0.0041 0.0091 0.1537 0.0007 0.0007
 [8,] 0.0283 0.0041 0.0007 0.0091 0.0091 0.0007 0.0007 0.0283 0.0007
 [9,] 0.0007 0.0041 0.0007 0.0041 0.0007 0.0007 0.0007 0.0007 0.0007

Selected shapes: 6 (BIC: 3610.8066)

Method: Least Squared Error

Mean squared error:
0.015710956518554

Available components:

 [1] "cluster"       "thetahat"      "rho_hat"       "normalized_LL"
 [5] "MSE"           "method"        "h"             "blockcluster" 
 [9] "BIC"           "s"             "details"       "initial"      
\end{Soutput}
\end{Schunk}
The resulting model uses 6 shapes, with a BIC value of 3610.8066. 
This indicates that similar blocks in the initial \code{nethist} fit are merged into six shapes. Each shape has a combined probability obtained by pooling the blocks, resulting in six distinct probability values in \code{\$thetahat}.
The corresponding MSE is 0.0157.

Additional fields can be accessed directly. The \code{\$blockcluster} field stores the clustering assignments used to construct the shapes.
\begin{Schunk}
\begin{Sinput}
R> hnh_bm_LSE$blockcluster
\end{Sinput}
\begin{Soutput}
K-means clustering with 6 clusters of sizes 12, 8, 7, 12, 2, 4

Cluster means:
          [,1]
1 0.0091397444
2 0.0040564904
3 0.0282953761
4 0.0007011218
5 0.0778985507
6 0.1536647041

Clustering vector:
 [1] 5 3 6 1 1 3 1 1 2 6 6 1 2 2 3 1 3 5 1 2 1 4 3 4 1 2 1 6 3 2 4 1 1
[34] 4 4 3 4 2 4 2 4 4 4 4 4

Within cluster sum of squares by cluster:
[1] 7.186458e-05 6.453970e-06 3.185062e-04 8.850521e-06 5.907372e-05
[6] 2.005092e-03
 (between_SS / total_SS =  97.2 %)

Available components:

[1] "cluster"      "centers"      "totss"        "withinss"    
[5] "tot.withinss" "betweenss"    "size"         "iter"        
[9] "ifault"      
\end{Soutput}
\end{Schunk}
The results show that the majority of blocks (32 out of 45) are assigned to the three lowest-probability shapes, whose means are less than 0.01. 
In this example, many blocks have probabilities close to zero and are therefore merged into the three low-probability shapes.

Like the \code{nethist} objects, the \code{hnethist} objects are visualized using the \code{plot} method as follows.
\begin{Schunk}
\begin{Sinput}
R> plot(hnh_bm_PLL) # Figure 3(c)
R> plot(hnh_bm_LSE) # Figure 3(d)
\end{Sinput}
\end{Schunk}
Figures~\ref{fig:nethist_borrowmoney}(c) and (d) display hybrid network histograms obtained with \code{method="PLL"} and \code{method="LSE"}, respectively. 
These plots show patterns similar to those in Figures~\ref{fig:nethist_borrowmoney}(a) and (b) because \code{hnethist} uses the \code{nethist} result as an initial estimate. 
The key difference is that \code{hnethist} merges blocks, such as (1,4), (2,2), (4,5), and (5,5) blocks in Figure~\ref{fig:nethist_borrowmoney}(c), to form a single shape, resulting in estimates with lower variance.

\begin{figure}[t!]
\centering
\begin{subfigure}[b]{0.32\textwidth}
  \centering
\includegraphics{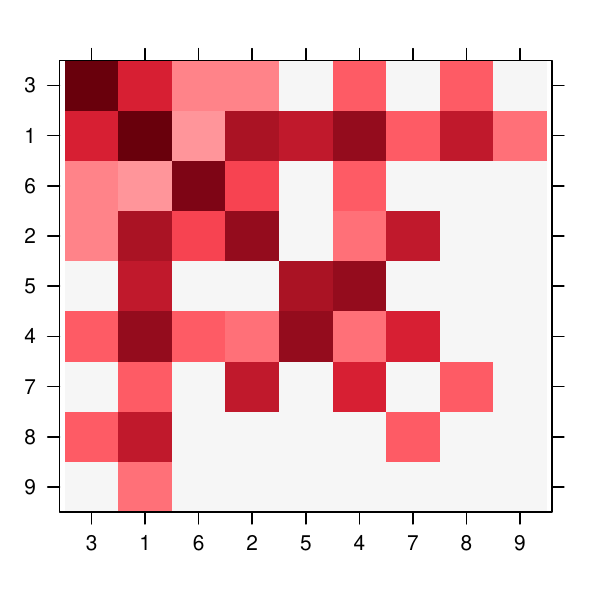}
  \caption{}
\end{subfigure}
\hfill
\begin{subfigure}[b]{0.36\textwidth}
  \centering
\includegraphics{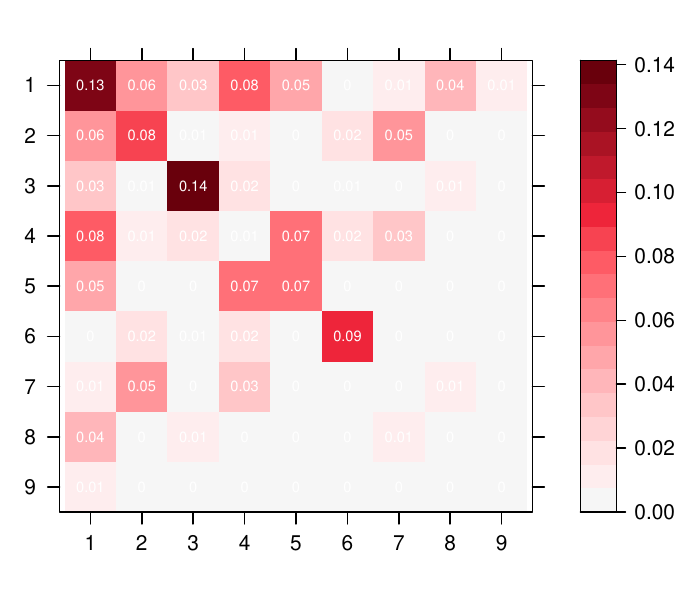}
  \caption{}
\end{subfigure}
\hfill
\begin{subfigure}[b]{0.28\textwidth}
  \centering
\includegraphics{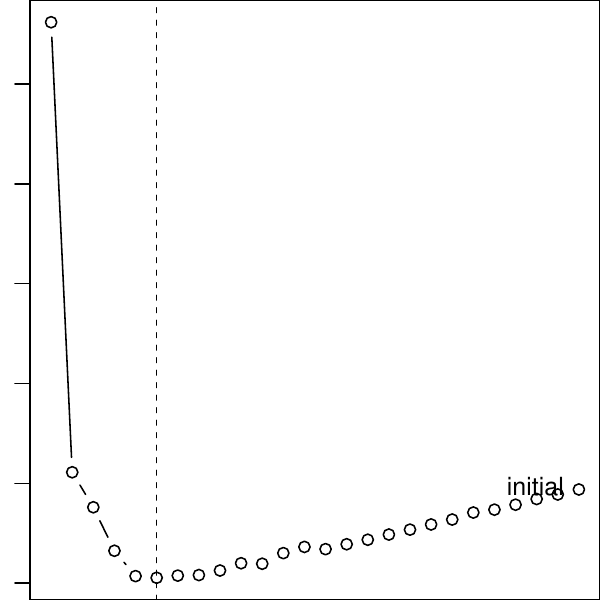}
  \caption{}
\end{subfigure}
\caption{\label{fig:plot_more} Network histogram results for
  the borrow money layer of the \code{IndianVil} dataset, including
  (a) permuted group labels, (b) estimated probability matrix with cell-wise annotations and a color key, and (c) BIC profile over the number of shapes, with the final model indicated by a vertical line.}
\end{figure}

Users may further customize the graphical outputs using the plotting arguments described in Section~\ref{subsec:vis_ftns}. The following code illustrates several customization options for the borrow money layer of the \code{IndianVil} dataset. 
\begin{Schunk}
\begin{Sinput}
R> plot(nh_bm_PLL, idx_order = order(diag(nh_bm_PLL$thetahat), 
+                             decreasing = TRUE))#Fib 4(a)
R> plot(nh_bm_PLL, type = "prob", prob = TRUE, colorkey = TRUE) #Fig 4(b)
R> plot(hnh_bm_LSE, type = "bic") #Fig 4(c)
\end{Sinput}
\end{Schunk}
Figure~\ref{fig:plot_more}(a) is obtained by reordering the group labels in Figure~\ref{fig:nethist_borrowmoney}(a) according to the diagonal entries of the estimated block probability matrix in decreasing order.
This reordering places blocks with higher connection probabilities toward the upper-left corner of the plot, allowing a clearer presentation of the block-wise density pattern.
Figure~\ref{fig:plot_more}(b) displays the estimated block probabilities stored in the \code{\$thetahat} field, with a color key shown on the right. Figure~\ref{fig:plot_more}(c) compares BIC values across models considered in the merging step. The dashed vertical line indicates the final model with 6 shapes. 

\subsection{Case 2: Multi-network histogram}\label{subsec:multilayer}

Rather than focusing only on the borrow money layer, jointly analyzing all 12 layers may better capture network characteristics among households in the village. This section demonstrates how to estimate scaled sets of graphons by applying \code{multinethist} to multilayer network observations.
The following code fits a multi-network histogram to all 12 layers of the \code{IndianVil} data.
\begin{Schunk}
\begin{Sinput}
R> set.seed(2026); mnh_IV <- multinethist(IndianVil)
\end{Sinput}
\end{Schunk}
The \code{multinethist} function selects a bandwidth of 23, which results in 10 groups.
\begin{Schunk}
\begin{Sinput}
R> mnh_IV$h #bandwidth from multinethist
\end{Sinput}
\begin{Soutput}
[1] 23
\end{Soutput}
\end{Schunk}
The \code{\$thetahat} field stores a 10$\times$10$\times$12 array of estimated connection probabilities, where each slice represents one of the 12 network layers. Layer-wise edge densities are available in the \code{\$rho\_hat} field as a vector of length 12.
\begin{Schunk}
\begin{Sinput}
R> dim(mnh_IV$thetahat)
\end{Sinput}
\begin{Soutput}
[1] 10 10 12
\end{Soutput}
\begin{Sinput}
R> length(mnh_IV$rho_hat)
\end{Sinput}
\begin{Soutput}
[1] 12
\end{Soutput}
\end{Schunk}

The data-driven bandwidth selection implemented in \code{multinethist} combines layer-wise features such as sparsity levels to obtain a common bandwidth. By leveraging connection patterns from denser layers, this method can provide a finer resolution even for sparser layers. To illustrate this, we compare the bandwidth values obtained from \code{multinethist} with 
those obtained by applying \code{nethist} to each layer. 
The layer-wise bandwidths from \code{nethist} vary substantially across 12 layers, ranging from 20 to 231, with a particularly large value in the temple company layer. 
\begin{Schunk}
\begin{Sinput}
R> single_h <- rep(0, 12); names(single_h) <- layer_spec
R> for(l in 1:12){
+    #To extract h quickly, set max_itr=1
+    single_h[l] <- nethist(IndianVil[,,l], 
+                           control = nethist_control(max_itr = 1))$h
+  }
R> single_h #layer-wise bandwidth from nethist
\end{Sinput}
\begin{Soutput}
  borrow money    give advice  help decision kero rice come 
            24             25             20             20 
  kero rice go     lend money          medic         nonrel 
            22             20             21             23 
           rel temple company     visit come       visit go 
            30            231             20             23 
\end{Soutput}
\end{Schunk}
The layer-wise bandwidths of sparser layers such as the temple company and rel layers are much larger than the bandwidth selected by \code{multinethist}.

As in the single-layer case, the multi-network histogram can be obtained using the \code{plot} method. The following code displays the 12 layer-specific plots in a 3-by-4 grid.
\begin{figure}[t!]
\centering
\includegraphics{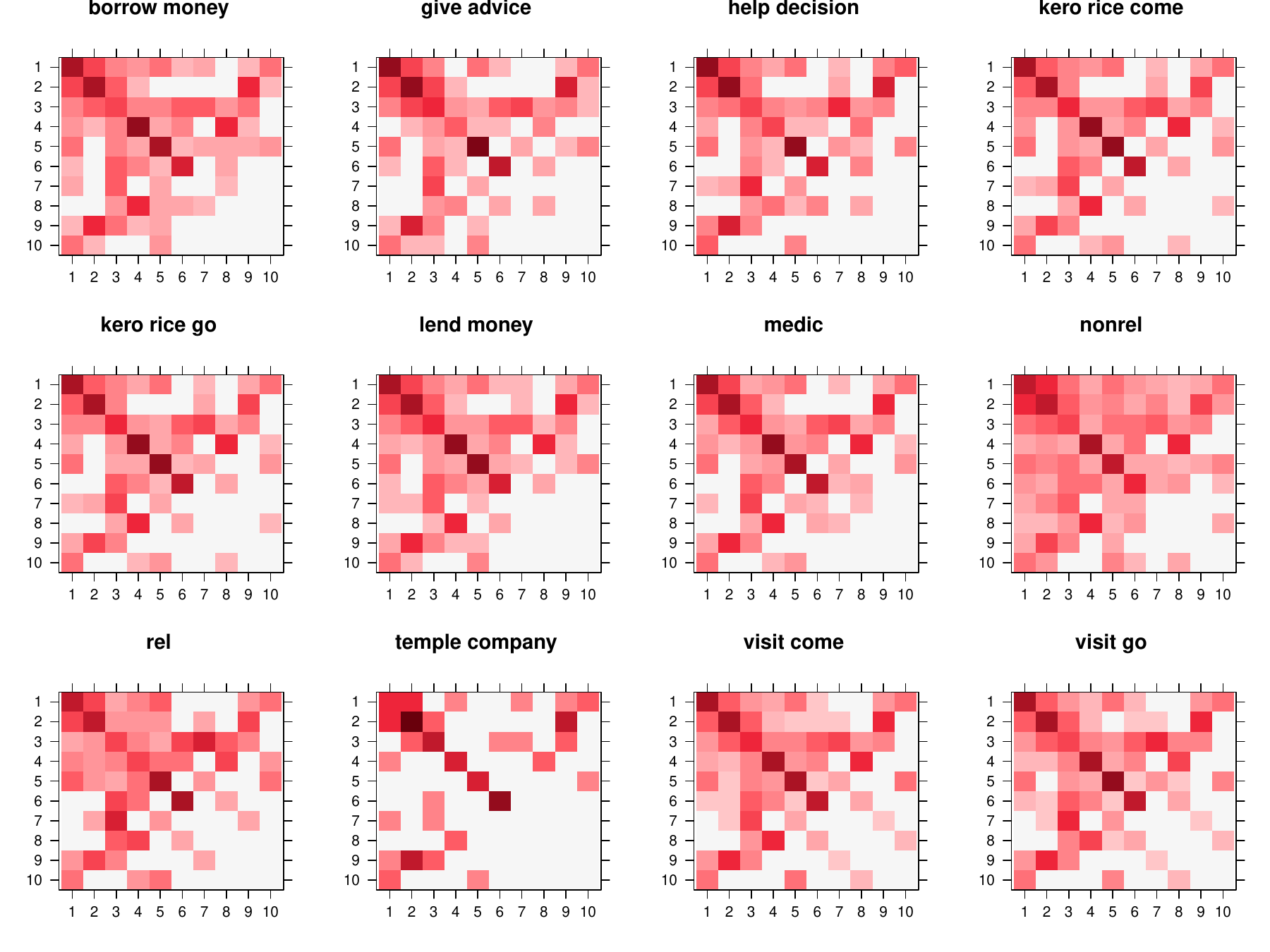}
\caption{\label{fig:multinethist} Multi-network histograms from \code{IndianVil} dataset. The plotted values are given by $(\hat{f}^{(\ell)})^{1/4}$ for each layer $\ell=1,\ldots,12$. All layers share identical group assignments.}
\end{figure}
\begin{Schunk}
\begin{Sinput}
R> plot(mnh_IV, layout = c(3,4), layer_titles = layer_spec)
\end{Sinput}
\end{Schunk}
Figure~\ref{fig:multinethist} shows the multi-network histograms of 12 layers. As all layers have a common \code{$cluster}, the group ordering is identical across layers. Most layers exhibit relatively large values along the diagonal blocks since the grouping procedure assigns vertices with similar connection patterns to the same group. 
In addition, layers representing closely related socio-economic relationships, such as (borrow money, lend money) and (visit come, visit go), exhibit similar network histograms.
Lastly, \code{multinethist} can provide higher-resolution histograms even for sparser layers by leveraging the group assignments obtained from all layers.
For example, in the temple company layer, \code{multinethist} produces a high-resolution representation that better captures the network structure, while the layer-wise bandwidth results in a single-block partition of vertices.

The \code{covariate_plot} can be used to examine group-wise characteristics when vertex covariates are available.
The following code illustrates its use with simulated covariates. We generate one numerical (\code{x\_num}) and one categorical covariate (\code{x\_cat}) that vary across groups.
\begin{Schunk}
\begin{Sinput}
R> set.seed(2026); grp <- mnh_IV$cluster
R> mu    <- rank(diag(mnh_IV$thetahat[,,1])) / max(grp) * 4
R> x_num <- rnorm(length(grp), mean = mu[grp], sd = 0.6)
R> brks  <- quantile(mu, c(1/3, 2/3))
R> x_cat <- factor(c("A","B","C")[findInterval(mu[grp] 
+                                + rnorm(length(grp), 0, 0.5), brks) + 1L])
\end{Sinput}
\end{Schunk}
\begin{figure}[t!]
\centering
\begin{tabular}{ll}
    \begin{minipage}{0.49\textwidth}
    \centering
\includegraphics{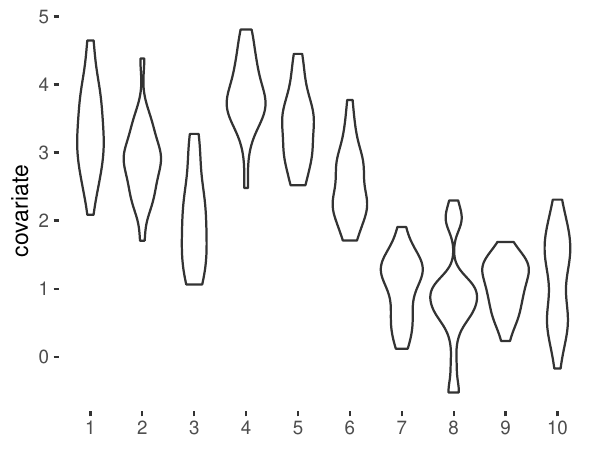}
    \par\vspace{0.3em}
    {\small (a) Numerical}
    \end{minipage}
    &
    \begin{minipage}{0.49\textwidth}
    \centering
\includegraphics{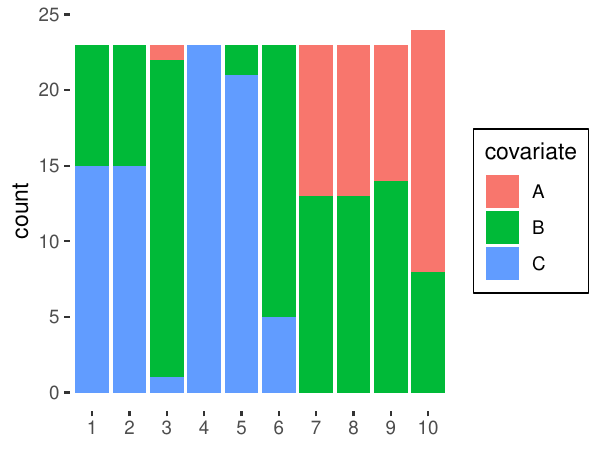}
    \par\vspace{0.3em}
    {\small (b) Categorical}
    \end{minipage}
\end{tabular}
\caption{\label{fig:covariate_plot} Covariate summary plot for the \code{IndianVil} dataset using simulated vertex covariates.}
\end{figure}
The following code produces covariate summary plots, as shown in Figure~\ref{fig:covariate_plot}(a) and (b) for the numerical and the categorical covariates, respectively.
\begin{Schunk}
\begin{Sinput}
R> covariate_plot(mnh_IV, x_num)
R> covariate_plot(mnh_IV, x_cat)
\end{Sinput}
\end{Schunk}

When layers are assumed to share an identical network structure, setting \code{common\_f = TRUE} enables homogeneous multi-network histogram fitting. This allows for a higher-resolution histogram with a lower MSE by averaging layer-specific network histograms. As an example, we assume that the borrow money, lend money, give advice, help decision, kero rice come, and kero rice go layers are generated from the same underlying graphon. The following code fits a homogeneous multi-network histogram to those layers.
\begin{Schunk}
\begin{Sinput}
R> layer_id <- 1:6
R> set.seed(2026) 
R> homo_mnh_IV <- multinethist(IndianVil[,,layer_id], common_f = TRUE)
\end{Sinput}
\end{Schunk}
The \code{plot} method produces the common histogram.
\begin{Schunk}
\begin{Sinput}
R> plot(homo_mnh_IV)
\end{Sinput}
\end{Schunk}
\begin{figure}[t!]
    \centering
\includegraphics{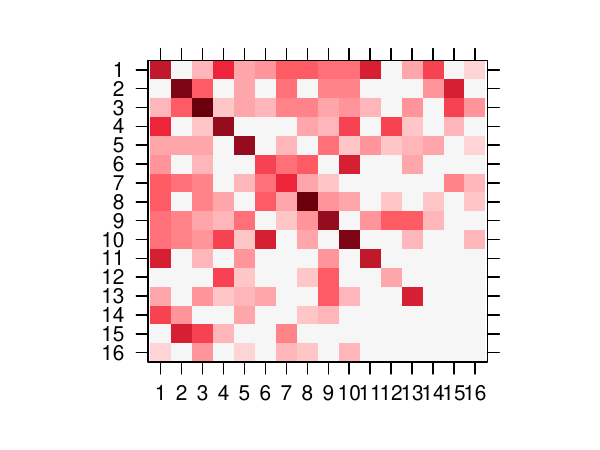}
\caption{\label{fig:homo_multinethist} homogeneous multi-network histogram fitted to six layers: the borrow money, lend money, give advice, help decision, kero rice come, and kero rice go layers. The displayed values show $\hat{f}^{1/4}$ to improve visualization.}
\end{figure}
Figure~\ref{fig:homo_multinethist} displays the homogeneous multi-network histogram with a bandwidth of 14, resulting in 15 groups of size 14 and one additional group of size 21 vertices. This bandwidth is smaller than that obtained from multi-network histograms in the earlier example, where graphons are not assumed to be identical across layers.
\begin{Schunk}
\begin{Sinput}
R> homo_mnh_IV$h
\end{Sinput}
\begin{Soutput}
[1] 14
\end{Soutput}
\begin{Sinput}
R> table(homo_mnh_IV$cluster)
\end{Sinput}
\begin{Soutput}
 1  2  3  4  5  6  7  8  9 10 11 12 13 14 15 16 
14 14 14 14 14 14 14 14 14 14 14 14 14 14 14 21 
\end{Soutput}
\end{Schunk}


\section{Discussion} \label{sec:summary}

This paper introduces the \code{nethist} \proglang{R} package,
a software implementation of network histogram methods for graphon estimation in both single-layer and multilayer networks. The current version supports the network histogram, the hybrid network histogram, and the multi-network histogram. Future implementations of the package may include additional estimators within the network histogram framework, such as graphon estimation with missing observations \citep{Gaucher2021Maximum} and covariate-assisted estimation via local linear smoothing \citep{chandna2022Local}. These extensions are applicable to practical settings where network data are partially observed or where vertex covariates provide information about latent variables.

We may also consider additional optimization strategies to improve the solution obtained by the greedy search procedure, which typically converges to a local optimum within reasonable computation time but may still be suboptimal. \citep[Chapter 3]{givensComputationalStatistics2012}.
The first direction enhances the greedy search procedure, for example, by incorporating multiple random starts and a best-of-$m$ rule.
The second direction explores alternative optimization algorithms beyond greedy search. For example, simulated annealing \citep{henderson2003Theorya} may help escape local optima by accepting non-improving swaps with a decreasing probability. From the user's perspective, new optimization algorithms or swap rules can be incorporated with minimal changes to \code{nethist_control}, as these modifications are largely handled within the internal implementation.


\section*{Computational details}

The results in this paper were obtained using
\proglang{R}~4.6.0 with the
\pkg{nethist}~1.0.0 package. \proglang{R} itself
and all packages used are available from the Comprehensive
\proglang{R} Archive Network (CRAN) at
\url{https://CRAN.R-project.org/}.

\section*{Acknowledgments}

\begin{leftbar}
\end{leftbar}


\bibliography{refs}

\end{document}